\documentclass[12pt,aps,prd,preprint,nofootinbib,superscriptaddress,nobalancelastpage]{revtex4-1}
\pdfoutput=1
\usepackage{amsfonts}
\usepackage{amssymb}
\usepackage{amsmath,amsthm,slashed,mathtools,mathrsfs,multirow}
\usepackage{graphicx}
\usepackage{array}
\usepackage{color}
\usepackage{ulem}
\usepackage{xspace}
\usepackage{verbatim}
\usepackage[usenames,dvipsnames]{xcolor}
\usepackage{subfigure}
\usepackage{verbatim}
\usepackage{dsfont}
\usepackage{braket}
\usepackage{mathtools}
\usepackage{adjustbox}
\usepackage{mleftright}
\usepackage{hyperref}\hypersetup{colorlinks,bookmarksopen,bookmarksnumbered,citecolor=blue,linkcolor=blue,pdfstartview=FitH,urlcolor=blue}

\DeclarePairedDelimiter\abs{\lvert}{\rvert}%
\newcommand{\theo}{{\rm th}}
\newcommand{\data}{{\rm d}}

\allowdisplaybreaks

\newcounter{mysubequation}[equation]

\definecolor{pink}{rgb}{1.,.2,.8}

\def\kth{Department of Physics, School of Engineering Sciences, KTH Royal Institute of Technology, AlbaNova University Center, 106 91 Stockholm, Sweden}

\def\ift{Instituto de F\'isica Te\'orica UAM/CSIC, Calle Nicol\'as Cabrera 13-15, Cantoblanco E-28049 Madrid, Spain}

\def\uam{Departamento  de  F\'{\i}sica Te\'{o}rica,  Universidad  Aut\'{o}noma  de  Madrid, Cantoblanco  E-28049  Madrid,  Spain}

\def\sissa{SISSA/INFN - Sezione di Trieste, Via Bonomea 265, I-34136 Trieste, Italy}

\def\icc{Institut de Ci\`encies del Cosmos, Universitat de Barcelona, Diagonal 647, E-08028 Barcelona, Spain}

\begin{document}

\preprint{FTUAM-18-7}
\preprint{IFT-UAM/CSIC-18-023}
\preprint{SISSA 09/2018/FISI}

\title{IceCube bounds on sterile neutrinos above 10 eV}

\author{Mattias Blennow}
\email{m.blennow@csic.es/emb@kth.se}
\affiliation{\ift}
\affiliation{\kth}

\author{Enrique Fernandez-Martinez}
\email{enrique.fernandez-martinez@uam.es}
\affiliation{\ift}
\affiliation{\uam}

\author{Julia Gehrlein}
\email{julia.gehrlein@uam.es}
\affiliation{\ift}
\affiliation{\uam}

\author{Josu Hernandez-Garcia}
\email{josu.hernandez@ts.infn.it}
\affiliation{\sissa}

\author{Jordi Salvado}
\email{jsalvado@icc.ub.edu}
\affiliation{\icc}

\begin{abstract}
\noindent
We study the capabilities of IceCube to search for sterile neutrinos with masses above 10~eV by analyzing its $\nu_\mu$ disappearance atmospheric neutrino sample. We find that IceCube is not only sensitive to the mixing of sterile neutrinos to muon neutrinos, but also to the more elusive mixing with tau neutrinos through matter effects. The currently released 1-year data shows a mild (around 2$\sigma$) preference for non-zero sterile mixing, which overlaps with the favoured region for the sterile neutrino interpretation of the ANITA upward shower. Although the null results from CHORUS and NOMAD on $\nu_\mu$ to $\nu_\tau$ oscillations in vacuum disfavour the hint from the IceCube 1-year data, the relevant oscillation channel and underlying physics are different. At the 99\% C.L.\ an upper bound is obtained instead that improves over the present Super-Kamiokande and DeepCore constraints in some parts of the parameter space. We also investigate the physics reach of the roughly 8 years of data that is already on tape as well as a forecast of 20 years data to probe the present hint or improve upon current constraints. 

\end{abstract}

\maketitle

\section{Introduction}
Over the last 20 years neutrino oscillations have been established as the explanation of the experimental evidence for neutrino flavour transitions~\cite{McDonald:2016ixn,Kajita:2016cak} with the mixing angles and mass squared differences measured to high accuracy (see Tab.~\ref{tab:exp_parameters} for recent global fit results of the mass and mixing parameters). The simplest extension of the Standard Model accommodating neutrino masses is the addition of sterile (right-handed) neutrinos to its content. The mass of these extra singlets, unlike all other fermions, would not be related to the Higgs mechanism and the electroweak scale due to their singlet nature. Therefore, it is vital to probe their existence experimentally at all possible scales. 

For instance, short baseline (SBL) experiments  like LSND~\cite{Aguilar:2001ty} and MiniBOONE~\cite{Aguilar-Arevalo:2012fmn, AguilarArevalo:2008rc} as well as reactor experiments combined with a recent reevaluation of their expected fluxes~\cite{Mention:2011rk,Huber:2011wv} and Gallium source experiments~\cite{Abdurashitov:2009tn, Kaether:2010ag,Giunti:2010zu} have reported oscillation results that are consistent with a mass squared difference with a possible fourth neutrino mass eigenstate of $\Delta m_{41}^2\approx 1~\text{eV}^2$, although this interpretation is in strong tension with other searches~\cite{Kopp:2013vaa,Gariazzo:2017fdh}. 
With an even higher mass, around the keV scale, sterile neutrinos are a viable dark matter candidate~\cite{Dodelson:1993je,Shi:1998km} that can be probed via their decay to light neutrinos and X-rays with both stringent constraints~\cite{Perez:2016tcq} and a possible hint at 3.5 keV~\cite{Bulbul:2014sua, Boyarsky:2014jta}, which could come from the decay of such neutrinos. At larger masses, sterile neutrinos would leave their imprint altering the kinematics of beta decays and meson decays and can also be probed for at beam dump and collider experiments~\cite{Atre:2009rg,Ruchayskiy:2011aa}. Even when beyond the reach of collider searches, sterile neutrino mixing can be tested indirectly via precision electroweak and flavour observables~\cite{Shrock:1980vy,Schechter:1980gr,Shrock:1980ct,Shrock:1981wq,Langacker:1988ur,Bilenky:1992wv,Nardi:1994iv,Tommasini:1995ii,Antusch:2006vwa,Antusch:2008tz,Biggio:2008in,Alonso:2012ji,Akhmedov:2013hec,Antusch:2014woa,Fernandez-Martinez:2015hxa,Abada:2015trh,Abada:2016awd,Fernandez-Martinez:2016lgt}.    

Through neutrino oscillation data, MINOS~\cite{Adamson:2011ku}, IceCube~\cite{TheIceCube:2016oqi}, SuperKamiokande~\cite{Abe:2014gda}, MiniBOONE~\cite{Cheng:2012yy}, and CDHS~\cite{Dydak:1983zq}, among others, have published limits on the sterile mixing parameters for mass squared differences in the range $\Delta m_{41}^2=0.01-10~\text{eV}^2$.
These results have been combined to global analyses in refs.~\cite{Kopp:2013vaa, Conrad:2012qt, Esmaili:2013vza, Collin:2016aqd,Dentler:2017tkw}. 

In this study we will consider larger sterile mass squared differences and investigate the sensitivity to the mixing of the sterile neutrinos to the $\mu$ and $\tau$ flavours of the presently released 1 year data, as well as forecasts for 8 years and 20 years, of atmospheric muon neutrino disappearance data at IceCube. In particular, we will study mass squared differences large enough for the sterile-neutrino-driven oscillations to be averaged out at IceCube energies ($\Delta m_{41}^2 \gtrsim 100~\text{eV}^2$) for atmospheric neutrinos traveling through the Earth ($L \lesssim 12000$~km).
Such mass squared differences are too big to explain the SBL anomalies but are compatible with the sterile neutrino interpretation~\cite{Cherry:2018rxj} of the upward directed cosmic ray shower observed by ANITA~\cite{Gorham:2016zah}. These mixings are however ruled out by cosmological constraints~\cite{Vincent:2014rja} and some non-standard effect suppressing the production of these sterile neutrinos in the early Universe would be necessary to reconcile the results~\cite{Vecchi:2016lty}. Our results apply for sterile neutrino masses $\Delta m_{41}^2 \gtrsim 100~\text{eV}^2$. Note that for sterile masses above 10 MeV stronger bounds on the active-heavy mixing with muon and tau neutrinos are present from laboratory experiments where the sterile neutrino could be detected directly~\cite{Atre:2009rg}.\footnote{Electron neutrino-sterile mixing can be constrained for even smaller mass squared differences via kink searches in $\beta$ spectra of certain isotopes and in neutrinoless double beta decay experiments~\cite{Atre:2009rg}.}
 
SuperKamiokande~\cite{Abe:2014gda} and DeepCore~\cite{Aartsen:2017bap} have already published constraints on the sterile mixing parameters in the averaged out regime. It should be noted that the energy threshold of SuperKamiokande is lower than the IceCube one and the averaged out regime for SuperKamiokande therefore starts at smaller mass squared differences ($\Delta m_{41}^2>10^{-1}~\text{eV}^2$). The same parameter space has also been probed by experiments like CHORUS~\cite{Eskut:2007rn} and NOMAD~\cite{Astier:2001yj} although, instead of analyzing the disappearance of atmospheric $\nu_\mu$ and the effect of the matter potential from neutral current interactions in presence of steriles, they searched for the appearance of $\nu_\tau$ in a $\nu_\mu$ beam through vacuum oscillations. We will compare our results to these current experimental bounds.

This paper is organized as follows: In Section~\ref{sec:sterile_mixing} we give an overview of the muon neutrino survival probability when the oscillations driven by the new mass eigenvalues are averaged out, in Section~\ref{sec:results} we analyse one year of though-going muon data in IceCube and give forecasts for the 8 years and 20 years sensitivities. Finally, in Section~\ref{sec:summary} we summarise our results and give our concluding remarks.

\begin{table}
\centering
\begin{tabular}{lcr} 
\hline
Parameter & best-fit ($\pm 1\sigma$) & $ 3\sigma$ range\\ 
\hline 
$\theta_{12}$ [$^{\circ}$] & $ 33.63^{+0.78}_{-0.75}$& $31.44\rightarrow 36.07$\\[0.5 pc]
$\theta_{13}$ [$^{\circ}$] & $ 8.52^{+0.15}_{-0.15}\oplus 8.55^{+0.14}_{-0.14}$& $8.07 \rightarrow 8.98$\\[0.5 pc]
$\theta_{23}$ [$^{\circ}$] & $ 48.7^{+1.4}_{-6.9}\oplus 49.1^{+1.2}_{-1.6}$ & $39.3 \rightarrow 52.4$\\[0.5 pc]
$\delta$ [$^{\circ}$] & $228^{+51}_{-33}\oplus 281^{+30}_{-33}$&$128\rightarrow 390$\\
\hline
$\Delta m_{21}^{2}$ [$10^{-5}$~eV$^2] $ & $7.40^{+0.21}_{-0.20}$ & $6.80\rightarrow 8.02$\\[0,5 pc]
$\Delta m_{31}^{2}$ [$10^{-3}$~eV$^2$]~(NO) &$2.515^{+0.035}_{-0.035}$&$2.408\rightarrow 2.621$\\[0,5 pc]
$\Delta m_{32}^{2}$ [$10^{-3}$~eV$^2$]~(IO) &$-2.483^{+0.034}_{-0.035}$&$-2.580\rightarrow -2.389$\\
\hline
\end{tabular}
\caption{The best-fit values and the 3$\sigma$ ranges for the mixing and mass parameters taken from ref.~\cite{Esteban:2016qun}.
There are two minima for $\theta_{13},~\theta_{23}$ and $\delta$. The first one corresponds to the normal mass ordering whereas
the second one corresponds to the inverted mass ordering. The 3$\sigma$ ranges are given for either ordering.
 }
\label{tab:exp_parameters}
\end{table}

\section{Sterile neutrino mixing}
\label{sec:sterile_mixing}

Upon the addition of several sterile neutrinos, the flavour eigenstates of the weak interactions $\ket{\nu_\alpha}$ ($\alpha=e,\ \mu,\ \tau,\ s_1,\ s_2,...$)  are related to the neutrino mass eigenstates $\ket{\nu_i}$, with masses $m_i$ ($i=1$, 2, 3, 4, 5,...) via the elements  $U_{\alpha i}$ of the lepton mixing matrix according to
\begin{align}
\ket{\nu_\alpha}=\sum_i U^*_{\alpha i}\ket{\nu_i}\,.
\end{align}
In general, the mixing matrix for $n$ neutrino flavours can be decomposed as the product of $n(n-1)/2$ rotations with mixing angles $\theta_{ij}$, with $(n-1)(n-2)/2$ physical phases $\delta_{ij}$. The usual parametrization is through a series of unitary rotations $V_{ij}$ in the $i$-$j$-plane given by
\begin{equation}
U = V_{3n} V_{2n} V_{1n} V_{3(n-1)} V_{2(n-1)} V_{1(n-1)} \cdots V_{34} V_{24} V_{14}\underbrace{ V_{23}V_{13}V_{12}}_{= U_0}\,,
\label{eq:order}
\end{equation}
with
\begin{equation}
(V_{ij})_{ab} = \begin{cases}
\cos(\theta_{ij}), & a = b \in \{i,j\} \\
\sin(\theta_{ij}) e^{i\delta_{ij}}, & a = i,\ b = j \\
-\sin(\theta_{ij}) e^{-i\delta_{ij}}, & a = j,\ b = i \\
1, & a = b \notin \{i,j\} \\
0, & \mbox{otherwise}
\end{cases}\,,
\end{equation}
and where $U_0$ has the usual PMNS matrix, $U_\nu$, as the upper left $3\times 3$ block. Note that we have not included rotations in the purely sterile sector, e.g., $V_{45}$, as such rotations are unphysical. Written in this fashion, the full mixing matrix takes the block form
\begin{equation}
U = \mathcal U U_0 = \begin{pmatrix}
1-\alpha & \Theta \\ X & Y
\end{pmatrix} \begin{pmatrix}
U_\nu & 0 \\ 0 & 1
\end{pmatrix}
= \begin{pmatrix}
(1-\alpha) U_\nu & \Theta \\
 X U_\nu & Y
\end{pmatrix}\,.
\end{equation}
Here, if the rotations are performed in the order given by Eq.~(\ref{eq:order}), $\alpha$ is a lower triangular matrix of the form~\cite{Xing:2007zj,Xing:2011ur,Escrihuela:2015wra,Li:2015oal,Blennow:2016jkn}
\begin{equation}
\alpha = \begin{pmatrix}
\alpha_{ee} & 0 & 0 \\ \alpha_{\mu e} & \alpha_{\mu\mu} & 0 \\ \alpha_{\tau e} & \alpha_{\tau\mu} & \alpha_{\tau\tau}
\end{pmatrix}\,,
\end{equation}
whose components to leading order in the active-heavy mixing elements are given by
\begin{equation}
\alpha_{\beta\gamma} \simeq \begin{cases}
\frac{1}{2}\sum_{i=4}^n \lvert U_{\beta i}\rvert^2, & \beta = \gamma \\
\sum_{i=4}^n U_{\beta i} U_{\gamma_i}^*, & \beta > \gamma \\
0, & \gamma > \beta
\end{cases}.
\end{equation}

The $\nu_\mu \to \nu_\mu$ oscillation probability, $P_{\mu\mu}$, will here be derived and discussed in the case where the active-heavy mixing angles are small, the corresponding mass squared differences are large enough for the oscillations to average out, and where the electron neutrinos do not participate in the oscillations\footnote{The oscillation of $\nu_\mu$ to $\nu_e$ at the energies and baselines that characterize the IceCube data are strongly suppressed. Indeed, $\theta_{13}$ is small and the solar mass squared difference $\Delta m^2_{21}$ is too small for the oscillations with $\theta_{12}$ to develop. Finally, $\theta_{14}$ has been tightly constrained by electron neutrino disappearance experiments~\cite{Kopp:2013vaa} and would also play a subleading role.} (i.e., $\Delta m_{21}^2 L/2E \ll 1$ and $\theta_{1i} = 0$). We do so by considering a basis that is rotated by $\mathcal U$ relative to the flavour basis. In this basis, the neutrino oscillation Hamiltonian in matter takes the form
\begin{equation}
\tilde H = \begin{pmatrix}
H_0 & 0 \\ 0 & H_1
\end{pmatrix} + V_\text{NC} \mathcal U^\dagger\begin{pmatrix} 1 & 0 \\ 0 & 0\end{pmatrix} \mathcal U\,,
\end{equation}
where $H_0$ is the standard Hamiltonian for $\mu$-$\tau$ oscillations in vacuum, $H_1$ is a diagonal matrix containing large entries, and $V_\text{NC} = \mp G_F N_n/\sqrt{2}$ (with the upper sign for neutrinos and the lower sign for anti-neutrinos). The upper left $2\times 2$ block describing the $\nu_\mu$-$\nu_\tau$ oscillations (not including the electron neutrino states) can be treated separately, leading to the $2\times 2$ effective Hamiltonian
\begin{align}
\tilde H_0 &= H_0 + V_\text{NC} (1-\alpha^\dagger)(1-\alpha) \simeq H_0 - V_\text{NC} (\alpha + \alpha^\dagger) \nonumber \\
&= \frac{\Delta m_{31}^2}{4E} \begin{pmatrix}
-\cos(2\theta_{23}) & \sin(2\theta_{23}) \\ \sin(2\theta_{23}) & \cos(2\theta_{23})
\end{pmatrix} - V_\text{NC} \begin{pmatrix}
2\alpha_{\mu\mu} & \alpha_{\tau\mu}^* \\ \alpha_{\tau\mu} & 2\alpha_{\tau\tau}
\end{pmatrix}\,,
\end{align}
where the $\simeq$ represents equality up to a matrix proportional to unity and to leading order in $\alpha$. This can be rewritten as 
\begin{equation}
\tilde H_0 \simeq \frac{\Delta_m}{2} \begin{pmatrix}
-\cos(2\theta_m) & \sin(2\theta_m)\lambda^* \\
\sin(2\theta_m)\lambda & \cos(2\theta_m)
\end{pmatrix},
\end{equation}
where
\begin{eqnarray}
\nonumber
\Delta_m^2 &=& \left[\frac{\Delta m_{31}^2}{2E} \cos(2\theta_{23}) + 2V_\text{NC}(\alpha_{\mu\mu}-\alpha_{\tau\tau})\right]^2
+ \left|\frac{\Delta m_{31}^2}{2E}\sin(2\theta_{23}) - 2V_\text{NC} \alpha_{\tau\mu}\right|^2\,, \\
\sin^2(2\theta_m) &=& \frac{1}{\Delta_m^2} \left|\frac{\Delta m_{31}^2}{2E}\sin(2\theta_{23}) - 2V_\text{NC} \alpha_{\tau\mu}\right|^2\,,
\label{eq:masssin}
\end{eqnarray}
and $\lambda$ is a phase factor of modulus one. Rotating back to the flavour basis, the muon neutrino survival probability is given by
\begin{equation}
P_{\mu\mu} = \left(1-\alpha_{\mu\mu}\right)^4 \left(1 - \sin^2(2\theta_m) \sin^2\left(\dfrac{\Delta_m L}{2}\right) \right) + \sum_{i=4}^n \lvert U_{\mu i}\rvert^4\,,
\label{eq:pmumu}
\end{equation}
where the last term is a constant leaking term~\cite{Fong:2016yyh}. Note that, except for the leaking term, all the sterile neutrino effects are encoded in the matrix $\alpha$, in particular in the elements $\alpha_{\mu \mu}$, $\alpha_{\tau \tau}$, and $\alpha_{\tau \mu}$, regardless of how many sterile neutrinos are considered as long as they are all in the averaged out regime~\cite{Blennow:2016jkn}. However, in our analysis of IceCube data we will allow a free normalization of the events, given the large uncertainties in the atmospheric neutrino fluxes, thus there will be no sensitivity to the normalization factor $(1-\alpha_{\mu\mu})^4$ nor to the leaking term, which does not depend on energy nor baseline. 

At leading order in $\alpha$, and neglecting $\Delta m^2_{31}$ whose effect is negligible at the energies of the IceCube data sample, the following probability is obtained
\begin{equation}
P_{\mu\mu} \simeq 1 - V^2_\text{NC} |\alpha_{\tau\mu}|^2 L^2\,,
\end{equation}
where the overall normalization has also been dropped since we allow a free normalization in the analysis. In order to ease the comparison with existing constraints from SuperKamiokande~\cite{Abe:2014gda} and DeepCore~\cite{Aartsen:2017bap} and to make use of the nuSQuIDS software~\cite{Delgado:2014kpa,nusquids} for numerical calculations without approximations, we will now particularize these expressions for the addition of a single sterile neutrino. With our given parametrization, we find that
\begin{equation}
U_{\mu4} = s_{24} e^{-i\delta_{24}}  \quad \text{and} \quad U_{\tau4} = c_{24} s_{34} \,, 
\end{equation}
so that
\begin{equation}
\alpha_{\mu\mu} = 1-c_{24} \simeq |U_{\mu4}|^2/2 , \quad \alpha_{\tau\tau} = 1-c_{34} \simeq |U_{\tau4}|^2/2, \quad
\alpha_{\tau\mu} = s_{24}s_{34} e^{i\delta_{24}} \simeq U_{\tau4} U_{\mu4}^*\,,
\end{equation}
and thus
\begin{equation}
P_{\mu\mu} \simeq 1 - V^2_\text{NC} |U_{\tau4}|^2 |U_{\mu4}|^2 L^2\,.
\label{eq:simpl}
\end{equation}
Therefore, the bounds will essentially follow a hyperbola in the $\abs{U_{\mu 4}}^2$-$\abs{U_{\tau 4}}^2$-plane.
\begin{figure}
\begin{center}
\includegraphics[scale=0.6]{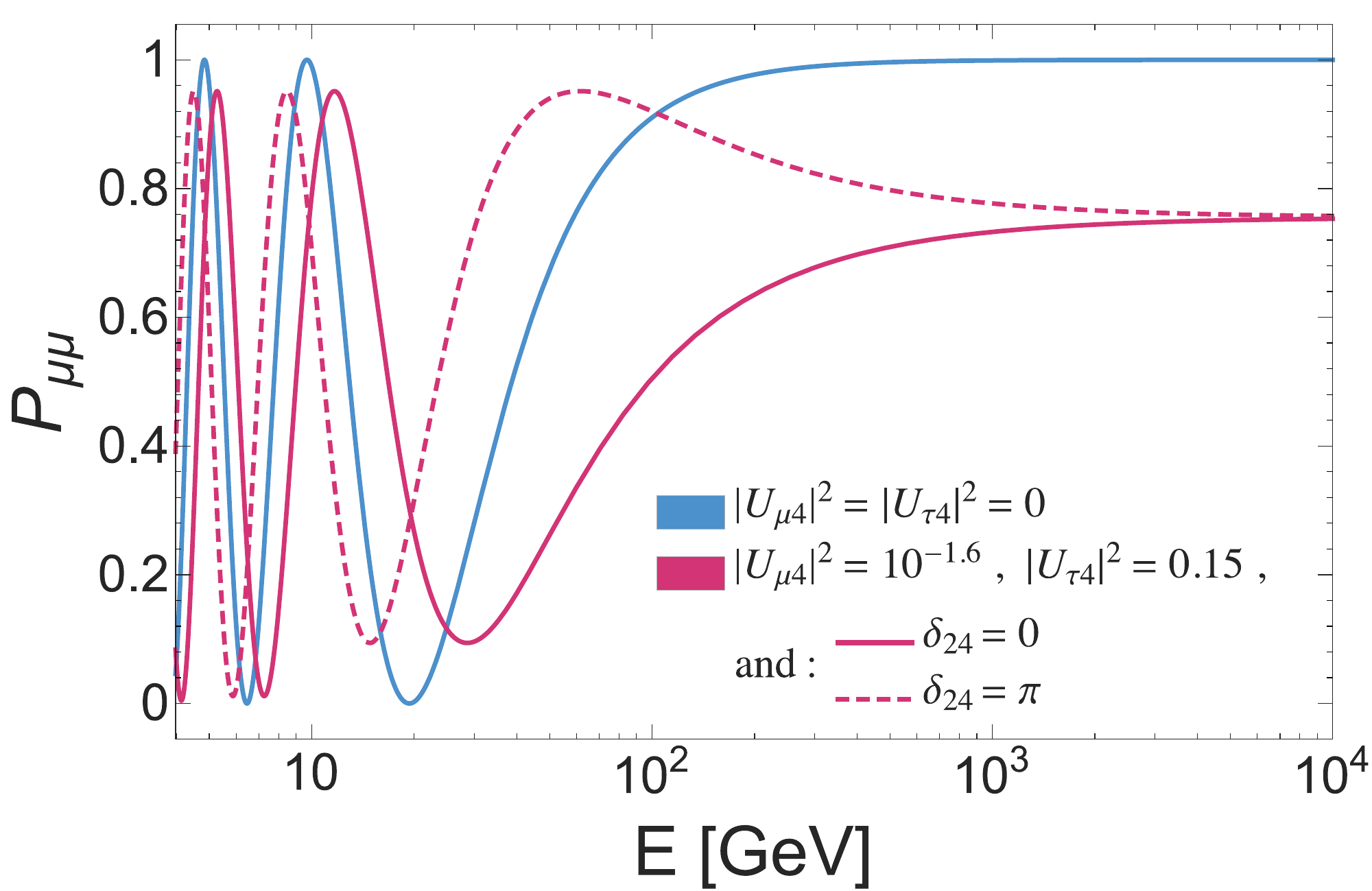}
\end{center}
\caption{\label{fig:analytical_prob} Muon neutrino survival probability using Eq.~\eqref{eq:pmumu} as a function of energy. The blue curve shows the oscillation probability without sterile mixing, while the magenta solid (dashed) curve shows the probability for $|U_{\mu4}|^2=10^{-1.6}$, $|U_{\tau4}|^2=0.15$, and $\delta_{24}=0$ ($\delta_{24}=\pi$). The baseline has been set to the diameter of the Earth.}
\end{figure}

Note that, in contrast to IceCube, for the SuperKamiokande and DeepCore energies the atmospheric oscillation driven by $\Delta m^2_{31}$ is relevant. Thus, the approximate Eq.~(\ref{eq:simpl}) is not valid and the sensitivity mainly stems from the interference between the standard and sterile oscillations in Eqs.~(\ref{eq:masssin}). Therefore, the phase of $\alpha_{\tau \mu}$, i.e., $\delta_{24}$ in the one extra sterile neutrino scenario, has an impact on the oscillation probability. Specifically, it can change the sign of the interference term between the atmospheric and the sterile terms in the expression for the energy and the mixing angle in matter. As an example of the impact of the phase, in Figure~\ref{fig:analytical_prob} the muon neutrino survival probability as a function of the energy for $|U_{\mu4}|^2=10^{-1.6},~|U_{\tau4}|^2=0.15$, $L = 1.2 \times 10^4$ km, and two different values of the phase, $\delta_{24}=0$ (solid line) and $\delta_{24}=\pi$ (dashed line) is shown. As comparison, the muon neutrino disappearance oscillation probability for zero sterile mixing is also shown. These values of the sterile matrix elements are at the border of the $90\%$ C.L. region of SuperKamiokande. The sign of the interference term can also be changed by changing the mass ordering (i.e., the sign of $\Delta m^2_{31}$) or by switching between neutrinos and antineutrinos (i.e., changing the sign of $V_\text{NC}$). However, neither IceCube nor SuperKamiokande or DeepCore can distinguish between neutrinos and antineutrinos so this dependence is diluted in their data. 

Conversely, experiments such as CHORUS and NOMAD explored the same parameter space but instead exploiting the $\nu_\mu$ to $\nu_\tau$ appearance channel with negligible matter effects leading to
\begin{equation}
P_{\mu\tau} \simeq 4 |U_{\tau4}|^2 |U_{\mu4}|^2 \sin^2 \left( \frac{\Delta m^2_{41} L}{4E}\right).
\label{eq:nomad}
\end{equation}

\section{Simulation and results}
\label{sec:results}

One year of high-energy through-going muons released by the IceCube collaboration~\cite{TheIceCube:2016oqi} for the last IceCube detector stage with 86 strings will be analyzed. The data sample consists of up-going track events so as to avoid the background from cosmic ray muons giving, after all cuts, a sample purity better than 99.9\%. Hence, the distances the signal neutrinos travel are of the order of $10^4$ km. The selected events have reconstructed energies between 400~GeV and 20~TeV and cosine of the reconstructed zenith angle between $-1$ and $0.2$. The sensitivity that a full 8-year IceCube sample would have as well as the prospects for an exposure equivalent to 20 years of IceCube data will also be forecasted. For our simulations, the neutrino flux computed with the analytic air shower code~\cite{Fedynitch:2015zma} using the cosmic ray flux from HondaGaisser model with Gaissser-Hillas H3a correction~\cite{Gaisser:2013bla} together with the hadronic model QGSJET II-04~\cite{Ostapchenko:2010vb} have been adopted. We have also verified that our results do not change significantly under the assumption of different fluxes, such as using the cosmic ray flux from the poly-gonato model~\cite{TerAntonyan:2000hh, Hoerandel:2002yg} or the Zatsepin-Sokolskaya~\cite{Zatsepin:2006ci} model  updated with measurements by PAMELA~\cite{Adriani:2011cu} together
with the hadronic model SIBYLL2.3, RC1, point-like~\cite{Fletcher:1994bd} or QGSJET II-04.

The propagation of the neutrinos was simulated using the nuSQuIDS software~\cite{Delgado:2014kpa,nusquids}, where the PREM profile~\cite{Dziewonski:1981xy} is implemented for the Earth matter density. Since we are interested in the averaged out regime our simulations were performed with a sterile mass squared difference of $\Delta m_{41}^2=10^3~\text{eV}^2$, but we have verified that changing this parameter does not alter the results as long as $\Delta m_{41}^2 \gtrsim 100~\text{eV}^2$ as expected. 
 
Since neutrino and antineutrino interactions cannot be distinguished on an event basis, the signal will contain both $\nu_\mu$ and $\bar{\nu}_\mu$ events.
After propagating the flux for every value of the sterile neutrino parameter, the Monte Carlo provided with the data releas~\cite{TheIceCube:2016oqi} has been used to compute the expected number of events $N_{\theo, i}$ in every bin of reconstructed zenith angle.

In order to obtain the expected significance of the bounds on the sterile mixing parameters,
we adopt a Poisson log-likelihood given by
\begin{align}
L=-\sum_{i} \left[N_{\theo, i}-N_{\data, i}+N_{\data, i}\log\left(\frac{N_{\data, i}}{N_{\theo, i}}\right)\right]\,,
\end{align}
where the $N_{\theo, i}$ and $N_{\data,i}$ are the predicted and observed number of events given a set of parameters in bin $i$, respectively, and the sum is taken over all the reconstructed zenith angle bins $i$.

The log-likelihood has been maximized for a number of nuisance parameters to include the effect of possible systematic errors. In particular, the uncertainty in the pion-kaon ratio of the initial flux ($\pi/k$),  the efficiency of the digital optical modules (DOMs), and the overall flux normalization have been considered. Since the observable is energy independent for large values of the sterile neutrino mass (see Eq.~\eqref{eq:simpl}), only one energy bin has been considered and the uncertainty in the energy spectrum slope has been neglected, while 40 bins for the reconstructed zenith angle have been adopted. 
For the pion-kaon ratio a Gaussian prior with $\sigma_{\pi/k}=0.05$ has been adopted and no prior for the DOM efficiency or the overall flux normalization has been assumed.
The standard oscillation parameters used in the simulations were set to their respective best-fit values from Tab.~\ref{tab:exp_parameters}. To find the confidence regions from the log-likelihood differences we assume that the prerequisites for Wilks' theorem~\cite{Wilks:1938dza} holds so that likelihood ratios can be directly converted to a confidence level.
\begin{figure}[t!]
 \includegraphics[width=0.48\columnwidth]{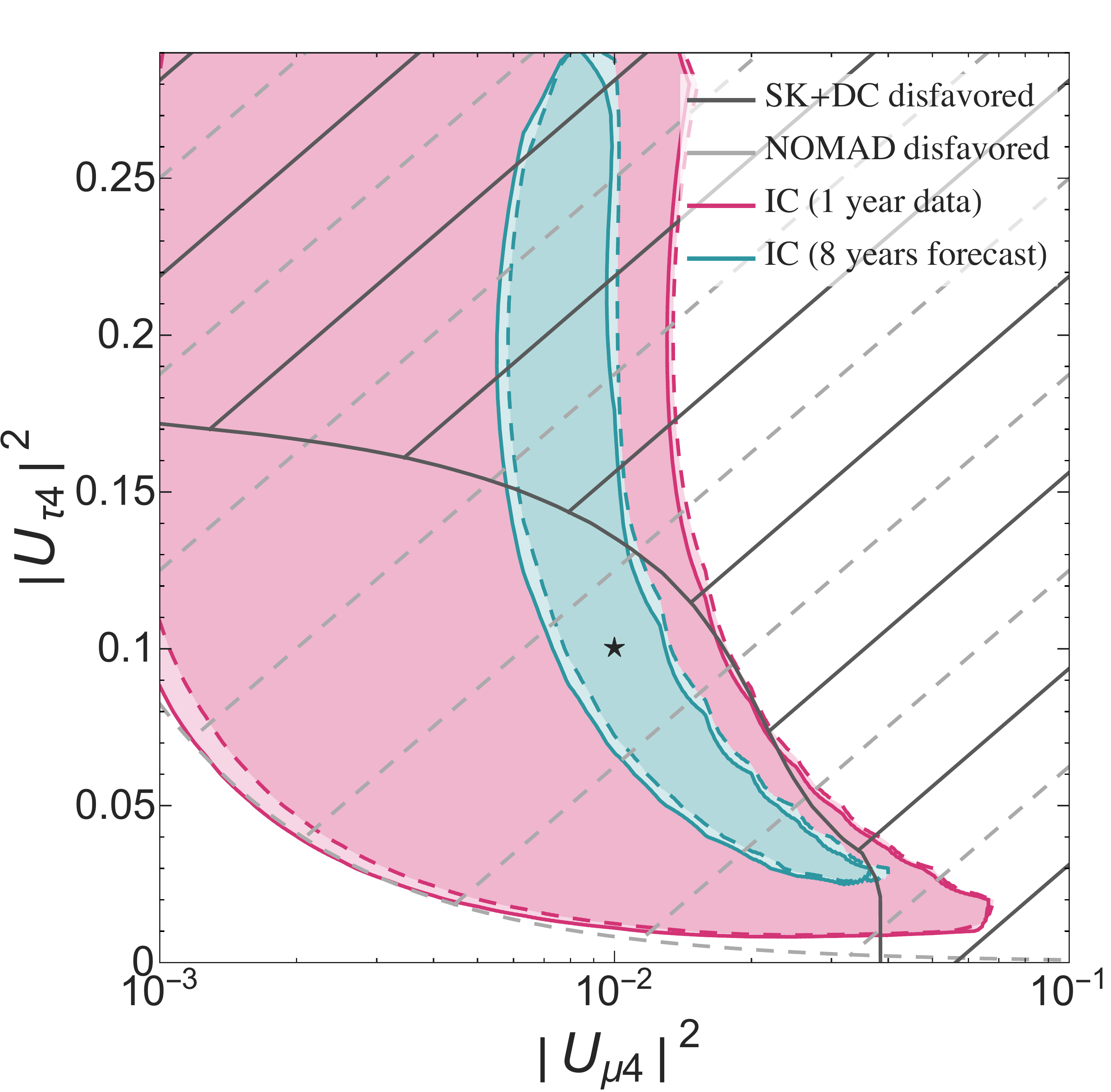} 
 \includegraphics[width=0.48\columnwidth]{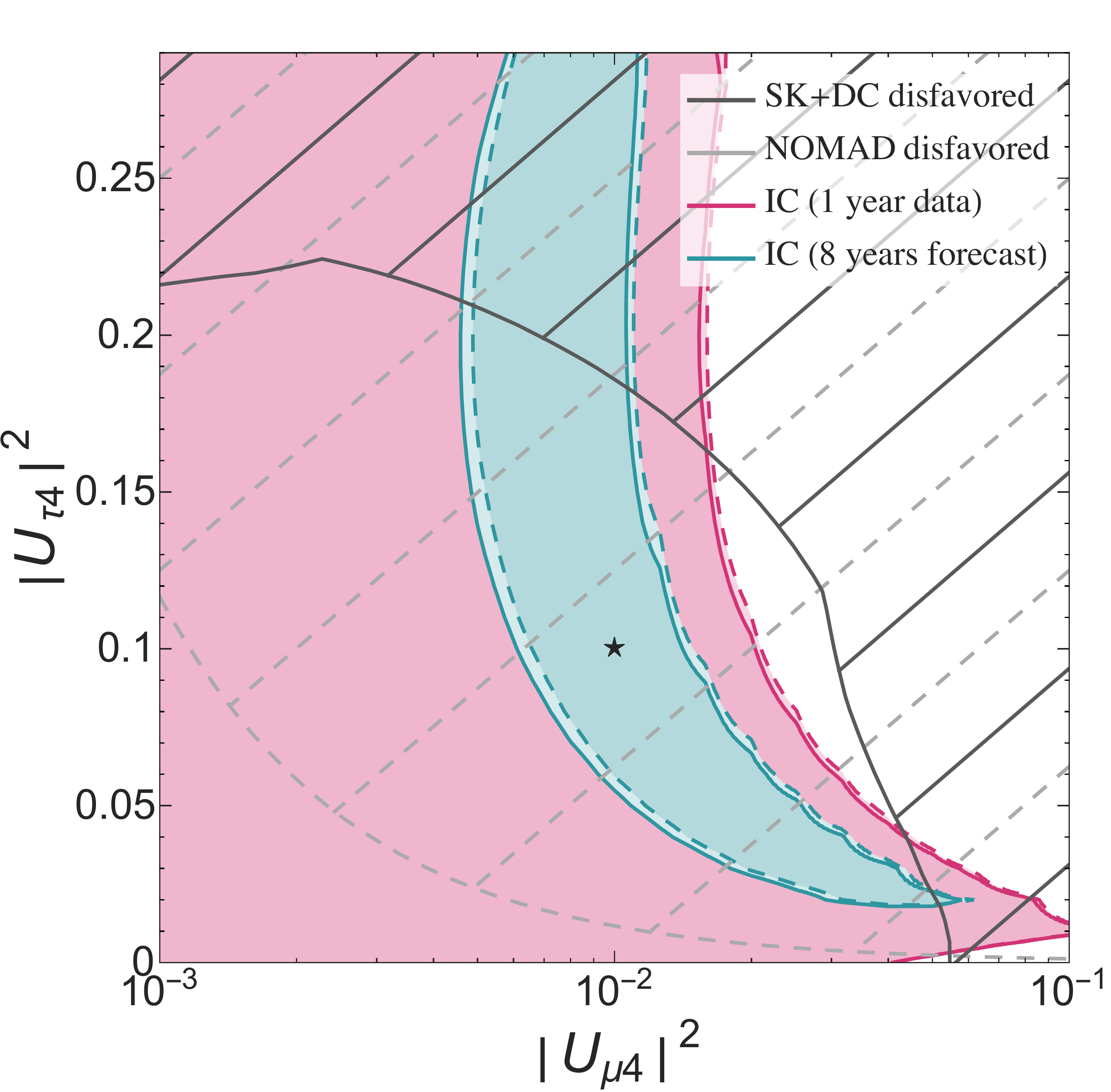}  
 \begin{center}
\caption{\label{fig:realdata_plot} The left (right) panel shows in pink the constraints at $90\%$ ($99\%$) C.L. for the sterile mixing elements from the released 1-year data. The cyan region shows at the same C.L.\ the forecast for 8 years of IceCube data assuming as true values $|U_{\mu4}|^2=10^{-2},~|U_{\tau4}|^2=0.1,~\delta_{24}=0$ (marked with a star).  The full (dashed) lines show the bounds for $\delta_{24} = 0$ ($\delta_{24} = \pi$). The solid (dashed) hatched regions are disfavoured by SuperKamiokande~\cite{Abe:2014gda} and DeepCore~\cite{Aartsen:2017bap} (NOMAD~\cite{Astier:2001yj}) data at the same C.L.}
\end{center}
\end{figure}

In the left panel of Figure~\ref{fig:realdata_plot}, the 90\% C.L. constraints (for 2 degrees of freedom) obtained for the public 1-year data (pink contours) in the $\abs{U_{\mu4}}^2$-$\abs{U_{\tau4}}^2$-plane is presented. The existing bounds from SuperKamiokande~\cite{Abe:2014gda} and DeepCore~\cite{Aartsen:2017bap} at the same C.L. are also shown for comparison by the hatched gray area. At 90\% C.L. present data prefer some degree of sterile mixing and we find that zero sterile mixing is disfavoured at 2.3$\sigma$ (1 degree of freedom\footnote{Note that if $U_{\mu 4}=0$, the $\nu_\mu$ survival oscillation probability is insensitive to $U_{\tau 4}$.}). The preference for non-zero sterile mixing is independent on the atmospheric sterile neutrino flux adopted in the analysis but its significance varies between 1.6 and 3.0~$\sigma$ with the different models tested. Given this preference for non-zero sterile mixing, the current constraints from IceCube do not improve upon the combined bounds from SuperKamiokande and DeepCore at 90\% C.L. In the right panel, the same information is shown at 99\% C.L. In this case, the present 1-year data gives an upper bound that already slightly improves upon the present SuperKamiokande and DeepCore constraints, ruling out the white region in the plot. 

The physics reach of an 8-year run of IceCube data if the present preference for sterile mixing is maintained is also shown in cyan. In particular, the present best-fit value of $\abs{U_{\mu4}}^2=10^{-2}$, $\abs{U_{\tau4}}^2=0.29$ lies in the already disfavoured region by DeepCore and SuperKamiokande. Due to the hyperbola-shaped degeneracy of the oscillation probability in the $\abs{U_{\mu4}}^2$-$\abs{U_{\tau4}}^2$-plane, there are values of the sterile oscillation parameters that provide an almost equally good fit without being in tension with the other $\nu_\mu$ disappearance present data. Remarkably, theses values of $U_{\tau4}$ are also compatible with the sterile neutrino interpretation~\cite{Cherry:2018rxj} of the upward directed cosmic ray shower observed by ANITA~\cite{Gorham:2016zah}. Indeed, the sterile neutrino interpretation of the ANITA results requires that the sterile neutrino mass is between $\sim 10^{2}$ and $\sim 10^{6}$~eV, which would also fall in the averaged out regime for IceCube studied here. However, all the parameter space preferred by IceCube at the 90\% C.L.\ is disfavoured by NOMAD~\cite{Astier:2001yj} with the same significance. Indeed, the null results in their $\nu_\tau$ search translates through Eq.~(\ref{eq:nomad}) into $\abs{U_{\mu4}}^2 \abs{U_{\tau4}}^2 < 8.3 \cdot 10^{-5}$ at the 90\% C.L.\ for $\Delta m^2_{41} \gtrsim 100$~eV$^2$. Nevertheless, the channel and underlying physics explored to obtain the bounds are very different in the two sets of experiments. While SuperKamiokande, DeepCore and IceCube analyze $\nu_\mu$ disappearance and the steriles are probed via their matter effects as shown in Eq.~(\ref{eq:simpl}), NOMAD and CHORUS searched for $\nu_\tau$ appearance essentially in vacuum through Eq.~(\ref{eq:nomad}). Thus, in presence of non-standard matter effects (also conceivably in the sterile sector) the two results could still be reconciled if a stronger tension should remain upon including more IceCube data. We therefore simulate 8 years of IceCube data assuming $\abs{U_{\mu4}}^2=10^{-2}$, $\abs{U_{\tau4}}^2=0.1$, and $\delta_{24}=0$ as the true oscillation parameters. As can be seen in Figure~\ref{fig:realdata_plot}, the expected confidence region region shrinks significantly with the additional statistics, while keeping its shape. In particular, if the values of the sterile neutrino mixing marked by the star were realized in nature, 8-years of IceCube data would disfavour no sterile mixing around the $5 \sigma$ level. 
\begin{figure}[t!]
 \includegraphics[width=0.48\columnwidth]{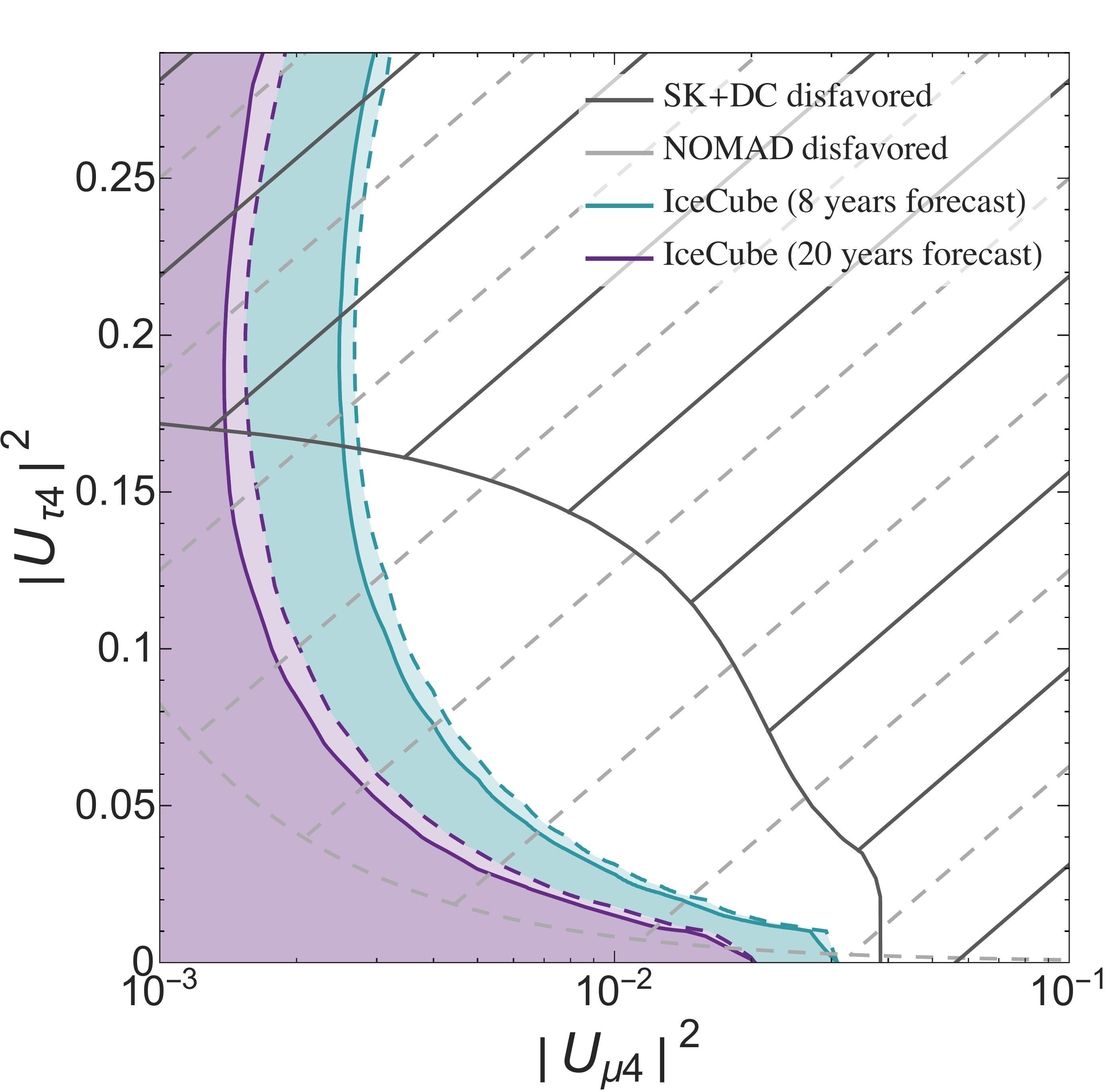} 
 \includegraphics[width=0.48\columnwidth]{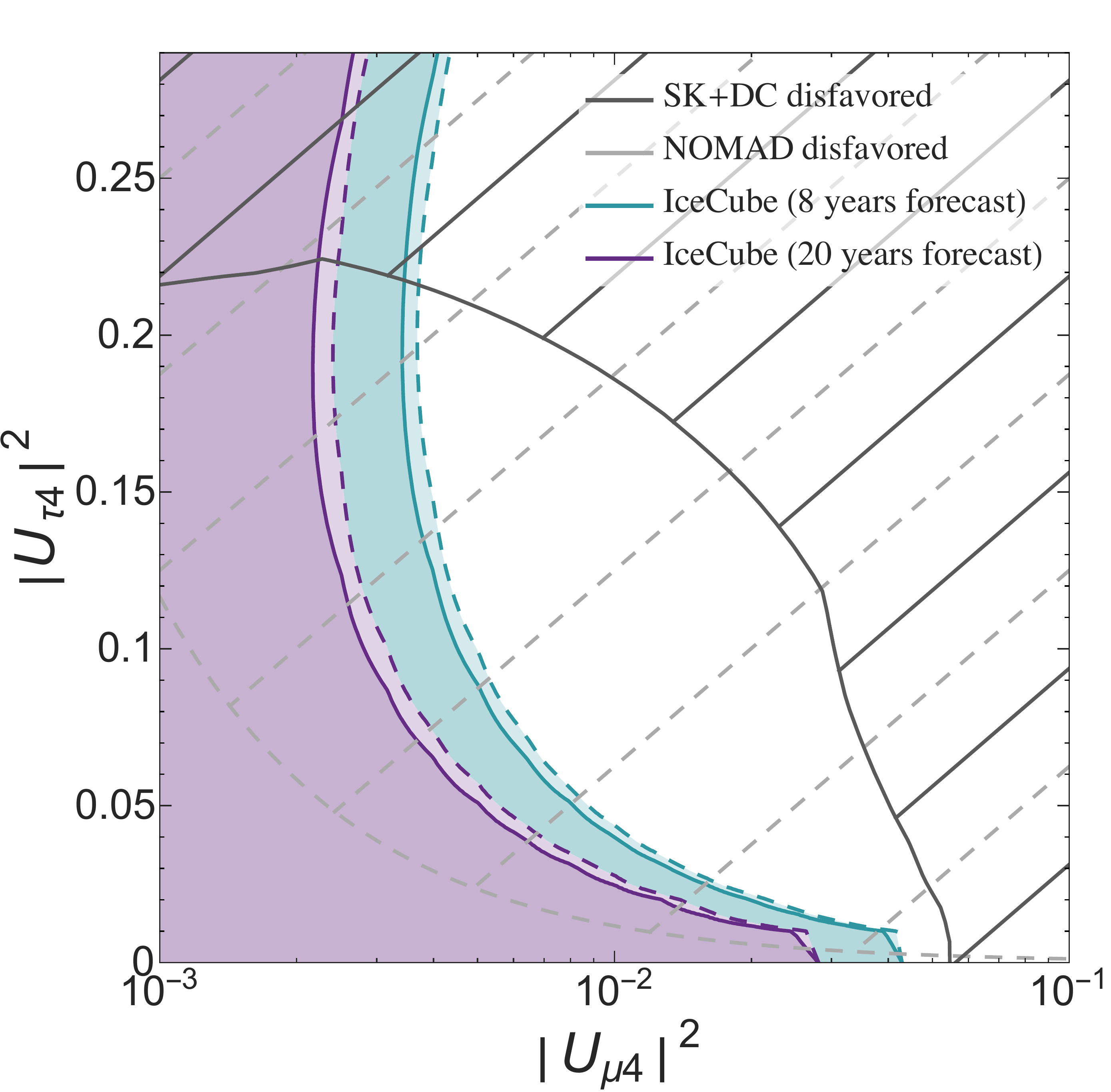}
 \begin{center}
\caption{\label{fig:forecast_plot} The left (right) panel shows the expected constraints in absence of sterile neutrino mixing at $90\%$ ($99\%$) C.L. for the sterile mixing elements from datasets composed of 8-year (cyan) or 20-year (purple) of IceCube data. The full (dashed) curves show the bounds for $\delta_{24} = 0$ ($\delta_{24} = \pi$). The solid (dashed) gray hatched regions are disfavoured by  SuperKamiokande~\cite{Abe:2014gda} and DeepCore~\cite{Aartsen:2017bap} (NOMAD~\cite{Astier:2001yj}) data at the same C.L.}
\end{center}
\end{figure}

The capability of larger IceCube samples to improve the present constraints on sterile mixing in absence of sterile neutrinos have been also studied.
In Figure~\ref{fig:forecast_plot}, the contours for $90\%$ (left panel) and $99\%$ C.L. (right panel) expected exclusion limits in the $\abs{U_{\mu4}}^2$-$\abs{U_{\tau4}}^2$-plane together with the existing bounds from SuperKamiokande and DeepCore are presented. The bound on $|U_{\mu4}|^2$ from 8 years of IceCube would improve over present constraints between a factor 1.3 for vanishing values of $|U_{\tau4}|^2$ to around an order of magnitude for $|U_{\tau4}|^2$ close to 0.1. Similarly, for $|U_{\mu4}|^2 \sim 10^{-2}$, the constraint on $|U_{\tau4}|^2$ would improve around a factor 5. In particular, the present best fit for non-zero sterile mixing would be excluded at high significance (more than $5 \sigma$) and most of the currently preferred parameter space at $90\%$ C.L. (pink area in the left panel of Figure~\ref{fig:realdata_plot}) disfavoured. Comparatively, increasing the statistics up to 20-year of IceCube data yields a more modest improvement in sensitivity. Remarkably, not even the 20-year scenario would improve over the present NOMAD limit of $\abs{U_{\mu4}}^2 \abs{U_{\tau4}}^2 < 8.3 \cdot 10^{-5}$ at the 90\% C.L. Nevertheless, we consider the two constraints complementary given the different physics probed by each of them.  

The effect of the CP-violating phase $\delta_{24}$ is also shown. In particular, the solid lines correspond to $\delta_{24}=0$ and the dashed lines to $\delta_{24}=\pi$. As can be seen, IceCube is not very sensitive to the sterile phase as oscillations due to the atmospheric mass squared difference at energies above 100~GeV do not have time to develop. Indeed, from Figure~\ref{fig:analytical_prob} the $\nu_\mu$ survival probability is essentially 1 in absence of sterile mixing for $E>100$~GeV.

\section{Summary and conclusions}
\label{sec:summary}

In this work we have presented the current constraints from the public 1-year IceCube data as well as the expectations of a full 8-year dataset and forecasts for 20 years worth of statistics to the mixing of sterile neutrinos with the $\mu$ and $\tau$ flavours. In particular, we concentrated for the first time on larger masses for the extra neutrinos ($\Delta m^2 > 100$~eV$^2$) than usually explored so that their oscillations are averaged out at IceCube. We find that the public 1-year IceCube data presents some preference for non-zero sterile mixing in the averaged out regime that would manifest via neutral-current-induced matter effects in the $\nu_\mu$ disappearance channel. In particular, values of the squared sterile mixing with the $\mu$ flavour of order $10^{-2}$ and with the $\tau$ between $10^{-1}$ and $10^{-2}$ are favoured at around $2 \sigma$ with respect to no sterile mixing. Interestingly, the large masses assumed in our analysis and the size of the preferred mixing with tau neutrinos correspond to the region of the parameter space that could also explain the upward directed cosmic ray shower observed by ANITA~\cite{Gorham:2016zah} with sterile neutrinos~\cite{Cherry:2018rxj}. These mixings are however in strong tension with cosmological constraints~\cite{Vincent:2014rja} and some non-standard effect suppressing the production of these sterile neutrinos in the early Universe would be necessary to reconcile these results~\cite{Vecchi:2016lty}. Moreover, these mixings are also in tension with present data from CHORUS~\cite{Eskut:2007rn} and NOMAD~\cite{Astier:2001yj} which, however, explore a different channel without matter effects. Thus, in presence of non-standard matter effects the two results could be potentially reconciled. 

We have also studied the sensitivity that 8 years of IceCube data, close to the data that should be presently available, would have and find that it would be sufficient to either confirm the present preference or exclude it with high significance (more than $5 \sigma$) and set stringent constraints improving around an order of magnitude over SuperKamiokande and DeepCore present bounds in some parts of the parameter space. Since sterile neutrinos at some mass scale are a general expectation of many extensions of the SM accounting for neutrino masses, it will be very interesting to explore this part of the parameter space with averaged out sterile neutrino oscillations using the full data sample collected by IceCube. 

\section*{Acknowledgments}

This work is supported in part by the European Union's Horizon 2020 research and innovation programme under the Marie Sklodowska-Curie grant agreements 674896-Elusives and 690575-InvisiblesPlus.  MB and EFM acknowledge support from the "Spanish Agencia Estatal de Investigaci\'on" (AEI) and the EU "Fondo Europeo de Desarrollo Regional" (FEDER) through the project FPA2016-78645-P; and the Spanish MINECO through the ``Ram\'on y Cajal'' programme and through the Centro de Excelencia Severo Ochoa Program under grant SEV-2012-0249. MB also acknowledges support from the G\"oran Gustafsson foundation. JS acknowledges support by MINECO grant FPA2016-76005-C2-1-P, Maria de Maetzu program grant MDM-2014-0367 of ICCUB and research grant 2017-SGR-929. JHG warmly thanks IFT of Madrid for its hospitality during part of this work, JG thanks the ITP, University of Heidelberg and EFM thanks the IPMU for its hospitality hosting him during the completion of this work. Finally, we acknowledge the use of the HPC-Hydra cluster at IFT. 

\providecommand{\href}[2]{#2}\begingroup\raggedright\endgroup

\end{document}